\documentclass[aps,twocolumn,superscriptaddress,floatfix,pt12,color]{revtex4}

\usepackage{graphicx}
\usepackage{epsfig}
\usepackage{amssymb}

\usepackage[dvips]{color}
\definecolor{Red}{rgb}{1.00, 0.00, 0.00}

\begin{document}
\newcommand{\be}{\begin{eqnarray}}
\newcommand{\ee}{\end{eqnarray}}
\newcommand\del{\partial}
\newcommand\nn{\nonumber}
\newcommand{\Tr}{{\rm Tr}}
\newcommand{\mat}{\left ( \begin{array}{cc}}
\newcommand{\emat}{\end{array} \right )}
\newcommand{\vect}{\left ( \begin{array}{c}}
\newcommand{\evect}{\end{array} \right )}
\newcommand{\tr}{\rm Tr}
\def\conj#1{{{#1}^{*}}}
\newcommand\hatmu{\hat{\mu}}
\newcommand\noi{\noindent}

\title{The Approach to the Thermodynamic Limit in Lattice QCD at $\mu\neq0$}

\author{K. Splittorff}
\affiliation{The Niels Bohr Institute, Blegdamsvej 17, DK-2100, Copenhagen
  {\O}, Denmark} 
\author{J.J.M. Verbaarschot$^{1,\, }$}
\affiliation{Niels Bohr International Academy, Blegdamsvej 17, DK-2100,
  Copenhagen {\O}, Denmark} 
\affiliation{Department of Physics and Astronomy, SUNY, Stony Brook,
 New York 11794, USA}

\date   {\today}
\begin  {abstract}
The expectation value of the complex phase factor of the fermion determinant 
is computed to leading order in the $p$-expansion of the chiral Lagrangian.  
The computation is valid for $\mu<m_\pi/2$ and determines
the dependence of the sign problem on the volume and on the geometric
shape of the volume.
In the thermodynamic limit with $ L_i \to \infty $ at fixed temperature
$1/L_0$, the average phase factor vanishes. In the low  temperature 
limit where $L_i/L_0$ is fixed  
as $L_i$ becomes large the average phase factor approaches one. The results 
for a finite volume compare well with lattice results
obtained by Allton  {\it et al}.. After taking appropriate limits, 
we reproduce previously derived results for the 
$\epsilon$-regime and for 1-dimensional QCD. The distribution of the phase
itself is also computed.

\end{abstract}

\maketitle
\newpage
 
\section{Introduction}

Numerical lattice QCD at nonzero baryon chemical potential is obstructed by
the sign problem:  At nonzero chemical potential, $\mu$, the phase factor of
the fermion determinant 
\be
e^{2i\theta} = \frac{{\det}(D +\mu\gamma_0+m)}
{\det(D+\mu\gamma_0+m)^*}
\ee
invalidates 
a direct application of Monte Carlo
methods. However, indirect methods have been devised to circumvent the sign
problem
\cite{Glasgow,KW,FOC,AFP,fodor1,fodor2,owe1,owe2,maria,gupta,Allton1,Allton2,Allton3,AANV,AFHL,schmidt}.
Since these approaches only apply when the average of 
the phase factor is close to unity it is of considerable interest to
understand when the fluctuations of the phase are mild and when they are
severe. Since the measurement of the average phase factor on the lattice is
plagued by the sign problem as well, it is imperative to 
understand the average phase factor analytically.   
   
In this paper we study the average phase factor analytically within chiral
perturbation theory. In particular, the approach to the thermodynamic limit
will be analyzed.

Despite the absence of baryons, chiral perturbation theory has proved a 
vital tool in understanding lattice simulations at nonzero baryon chemical 
potential. The generating functional for the eigenvalue density of the QCD
Dirac operator includes quarks with the opposite sign of the chemical
potential \cite{misha} which therefore couple to the third 
component of isospin. Using this fact, the exact 
quenched \cite{SplitVerb2} and unquenched \cite{O,AOSV} microscopic spectral 
density of the QCD Dirac operator were derived from a chiral
Lagrangian.  This result revealed that at nonzero chemical 
potential,
the chiral condensate is related to                
the spectral density by a mechanism
that is different from the Banks Casher relation \cite{BC}:
The discontinuity in the chiral condensate is
due to complex oscillations on the microscopic scale \cite{OSV} 
in a macroscopic region of the complex eigenvalue plane.

The microscopic limit is also known as the $\epsilon$-domain of
QCD. Microscopic results for QCD can equally well be derived by means of 
chiral random matrix theory \cite{VS,V}. 
This has the advantage that  one may employ powerful 
random matrix methods such as 
orthogonal polynomials \cite{Mehta,Fyodorov,Bergere,Akemann} 
the replica trick \cite{Edwards}   or the sumpersymmetric
method \cite{Efetov}. 
For example, the unquenched microscopic spectral
density at nonzero chemical potential 
was first derived by means of random matrix theory
\cite{O}, whereas the quenched spectral density at nonzero
chemical potential was first obtained by means of the replica trick
in combination with the Toda lattice equation \cite{SplitVerb2}.     
    
The recent computation of the average phase factor in the microscopic domain 
\cite{exp2ith-letter,phase-long} shows that the average
phase factor is suppressed exponentially with the volume when $\mu>m_\pi/2$. 
For such values of the chemical potential the quark mass is inside the cloud of
eigenvalues of the Dirac operator and numerical lattice QCD simulations
become exceedingly difficult. For smaller values of the chemical potential,
the quark mass is outside the two dimensional domain of the eigenvalues, and
the sign problem is less severe.

In this paper we examine the character of 
the sign problem in the region $\mu<m_\pi/2$ and temperatures such that the
use of chiral perturbation theory can be justified. With $\mu<m_\pi/2$ it was 
found in \cite{exp2ith-letter,phase-long} that the average phase factor remains
nonzero in the microscopic limit where $\mu F_\pi \sqrt{V}$ is held fixed as
the volume is taken to infinity.  For $\mu F_\pi \sqrt{V}\gg 1$
the large volume asymptotic limit of the microscopic prediction is simply  
given by
\be
\langle e^{2i\theta}\rangle_{N_f} = (1-\frac{4\mu^2}{m_\pi^2})^{N_f+1} \quad
 \qquad
\mu<m_\pi/2.
\label{phase-hatmu-large}
\ee 
(The quenched and the phase quenched average of the phase factor give 
identical predictions in this limit. Both are obtained by setting $N_f=0$ in
the equation above.)  
This result suggest that unquenched lattice simulations in this domain are
feasible.

Here we examine whether the average phase factor remains nonzero
for $\mu<m_\pi/2$ when we relax the microscopic constraints 
and approach 
the thermodynamic limit at fixed chemical potential. 
In order to do so we compute the average phase 
factor using the $p$-expansion of chiral perturbation theory where 
\be
p \sim 1/L, \quad m_\pi \sim  1/L, \quad \mu \sim 1/L , \quad T \sim 1/L,
\ee
and work to leading (one-loop) order. The previous calculation of the average
phase factor was worked out \cite{exp2ith-letter,phase-long} in the 
microscopic domain where $m_\pi^2F_\pi^2 \sim 1/V$ and $\mu^2F_\pi^2 \sim
1/V$ as the volume is taken to infinity. The new one-loop computation presented
here includes the effect generated by the nonzero momentum modes of the
Goldstone bosons.  
We keep explicitly the dependence on volume $V=L_i^3L_0$ 
and the ratios $L_0/L_i$
in order 
to study the approach to the thermodynamic limit. The new result bridges
the gap
between the microscopic prediction \cite{exp2ith-letter,phase-long} and the
parameter range typically used in lattice gauge theories. This allows us to
compare the one-loop result for the average phase factor to lattice results
by  Allton  {\it et al.}. Below the pseudo-critical temperature for chiral
symmetry restoration the
lattice results are in remarkably good agreement with the analytical 
predictions.     

The distribution of the phase itself (rather than the phase factor) also
follows from the one-loop computation. We give the explicit form of the
distribution of the phase  to one-loop order in chiral perturbation theory. 

The paper is organized as follows. In the next section
we present the general setup for computing 
the one-loop result for the average phase factor within chiral perturbation
theory. The explicit one-loop result is derived
 in section \ref{sec:G-G}.
This expression is evaluated numerically in section \ref{sec:num} and the
comparison to    
lattice data is made in section \ref{sec:lat}. The effect
of a finite box on the average phase factor is further discussed 
in section \ref{sec:resonance}. Section \ref{sec:theta-dist} contains the
discussion of the distribution of the phase. We end with concluding remarks in
section \ref{sec:concl}.

\section{The average phase factor to leading order} 
\label{sec:average-general}

The average phase factor in the full theory is the ratio of  two
partition functions. A partition function with an extra fermionic quark as
well as a conjugate bosonic quark divided by the usual QCD  partition
function 
\be
\label{exp2ith-quarks}
&& \langle e^{2i\theta}\rangle_{N_f} = \\
&&\frac{\langle{\det}^{N_f+1}(D+\mu\gamma_0+m)
        /\det(D-\mu\gamma_0+m)\rangle}
       {\langle\det(D+\mu\gamma_0+m)^{N_f}\rangle} .\nn
\ee
Here we used that conjugate quarks correspond to ordinary quarks with the
opposite sign of the chemical potential \cite{AKW}. 
With the usual setup of leading order chiral perturbation theory, see for
example \cite{HL,STV}, the free energy is a sum of contributions from each of
the Goldstone bosons.   
The contributions to the numerator of pions with no isospin charge  
cancel against contributions from the denominator. This
leaves   us with
\be
\langle e^{2i\theta}\rangle_{N_f} & = & e^{(N_f+1)(G_0(\mu=0)-G_0(\mu))},
\label{phase-G}
\ee
where each $G_0$ includes the contribution of two oppositely charged Goldstone 
modes. In order to get the combinatorics right,
notice that the inverse determinant in (\ref{exp2ith-quarks}) represents a
conjugate bosonic quark. The charged Goldstone bosons 
contain this bosonic quark in addition to 
one of the $N_f+1$ fermionic quarks and are thus fermionic in nature
resulting in an additional minus sign from the fermionic loop.

Because of the sign problem one often studies averages in the phase quenched
theory  where the phase of the fermion is ignored
\be
Z_{1+1^*} = \langle |\det(D+\mu\gamma_0+m)|^2 \rangle.
\ee   
The average phase factor in the phase quenched theory is defined by
\be
 \langle e^{2i\theta}\rangle_{1+1^*}
=\frac{ \langle {\det}^2(D+\mu \gamma_0 +m)\rangle}
{\langle |\det(D+\mu \gamma_0 +m)|^2\rangle} 
=\frac{Z_{N_f=2}}{Z_{1+1^*}}.
\ee
To  leading order in chiral perturbation theory this ratio is given by
\be
\langle e^{2i\theta}\rangle_{1+1^*}=e^{G_0(\mu=0)-G_0(\mu)}.
\label{deltaGpq}
\ee
Notice that this result coincides with the one-loop result for
 the quenched theory  obtained from
(\ref{phase-G}) by setting $N_f=0$ as well as with 
the result for partially
quenched computations with dynamical quarks at zero chemical potential.

In summary, to determine the average phase factor to one-loop order
all that is required is the difference between $G_0(\mu)$ and $G_0(\mu=0)$.
 We will derive this
difference in section \ref{sec:G-G}.

\section{Evaluation of $G_0(\mu)-G_0(\mu=0)$} 
\label{sec:G-G}

As we are particularly interested in the approach to the thermodynamic limit
we will consider a finite system with volume $V=L_i^3L_0$. Hasenfratz
and Leutwyler \cite{HL} worked out $G_0$ for $\mu=0$. This calculation was
generalized to nonzero chemical potential in  
\cite{phase-long}. For completeness, we repeat the
main steps of the computation below.

Consider a single charged Goldstone boson with charge $2$ in a box
$V=L_i^3L_0$. The one-loop contribution to the partition function is given by 
\be
e^{G_0(\mu)/2} \equiv \exp[-\frac 12 \sum_{p_{k\, \alpha}}
\log(\vec p^2_k +m^2_\pi +(p_{k\, 0}-2i \mu)^2) ],
\ee
where
\be
p_{k\,\alpha} = \frac {2\pi k_\alpha}{L_\alpha}, 
\qquad k_\alpha \quad {\rm integer}.
\ee
While $G_0(\mu)$ is divergent the difference between $G_0(\mu)$ and
$G_0(\mu)$ for $V=\infty$ is finite, that is 
\be
G_0(\mu) = G_0(\mu)|_{V=\infty} + g_0(\mu)
\ee
with $g_0(\mu)$ finite. The $p_0$-integration contour in the
second term of the difference
\begin{widetext}
\be
G_0(\mu)|_{V \to \infty} - 
G_0(\mu=0)|_{V \to \infty} \nn
=\int p^{d-1} dp dp_0 [\log(\vec p^2 +m_\pi^2 + p_0^2) 
-\log(\vec p^2 +m_\pi^2 +(p_0-2i\mu)^2)]
\ee
can be shifted by $2i\mu$ if there are no obstructions from singularities.
This is the case if $2\mu <m_\pi$, so that the $\mu$-dependence resides
entirely in $g_0(\mu)$. 
The difference of the free energies in (\ref{phase-G}) and (\ref{deltaGpq})
is thus given by the difference of the finite parts
\be
G_0(\mu)-G_0(\mu=0)=g_0(\mu)-g_0(\mu=0).
\label{finite}
\ee
This also shows that the average phase factor does not depend on the
ultra-violet cutoff. The infrared nature of the average phase factor 
has been verified on the lattice \cite{SplSve}.

After several manipulations  including Poisson resummation and Jacobi's
imaginary transformation (the steps are given in detail in \cite{phase-long})
we find two equivalent representations of $g_0(\mu)$ 
both valid for $\mu<m_\pi/2$,    
\be
g_0(\mu)
&=&
\int_0^\infty \frac{d\lambda}{\lambda^3}
e^{-m^2_\pi L^2\lambda /4\pi }
(\prod_{\alpha=0}^3 {\sum_{l_\alpha}} e^{-2\mu l_0 L_0\delta_{\alpha 0}} 
e^{ -\pi \frac{l_\alpha^2 L_\alpha^2}{\lambda L^2}} -1).
\ee
and
\be
\label{g0mu}
g_0(\mu)
=
\int_0^1 \frac{d\lambda}{\lambda^3}
e^{-m^2_\pi L^2\lambda /4\pi }(\prod_{\alpha=0}^3
{\sum_{l_\alpha}} e^{-2\mu l_0 L_0\delta_{\alpha 0}} 
e^{ -\pi \frac{l_\alpha^2 L_\alpha^2}{\lambda L^2}}-1) \nn\\
+ \int_0^1 \frac{d\lambda}{\lambda}
 e^{\frac{\mu^2L^2 } 
{\pi\lambda }}
e^{-m^2_\pi L^2/(4\pi \lambda  )}(\prod_{\alpha=0}^3 
{\sum_{l_\alpha}} e^{-2 i \mu l_0 \frac{ L^2}{ L_0 \lambda} \delta_{\alpha0}} 
e^{ -\pi l_\alpha^2 \frac{ L^2}{ L_\alpha^2 \lambda}}-1)  \nn\\
+\int_1^\infty \frac{d\lambda}{\lambda}
e^{\frac{\mu^2 L^2  \lambda } {\pi}}
e^{-m^2_\pi L^2\lambda /4\pi } -
 \int_1^\infty \frac{d\lambda}{\lambda^3}
e^{-m^2_\pi L^2\lambda /4\pi } ,
\ee
where $l_\alpha$ runs over all integers and we have introduced the length
\be
L \equiv (L_0L_i^3)^{1/4}.
\ee
The first representation has the advantage that
the chemical potential explicitly appears in the combination $\mu L_0$. (Note
that $L_0$ also appears in other places.). The second representation 
can be used to expand the result
in a power series in $m_\pi^2$ and $\mu^2$. It also may be 
preferable  for  numerical evaluations.

\begin{figure}[t!]
  \unitlength1.0cm
  \epsfig{file=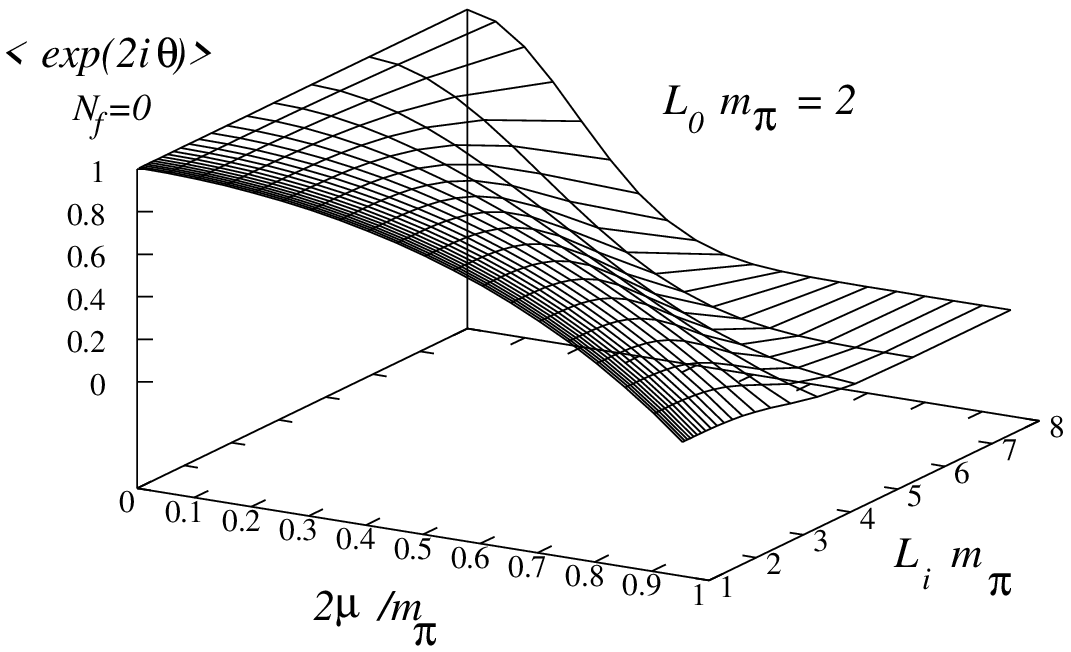,clip=,width=7cm}
 \epsfig{file=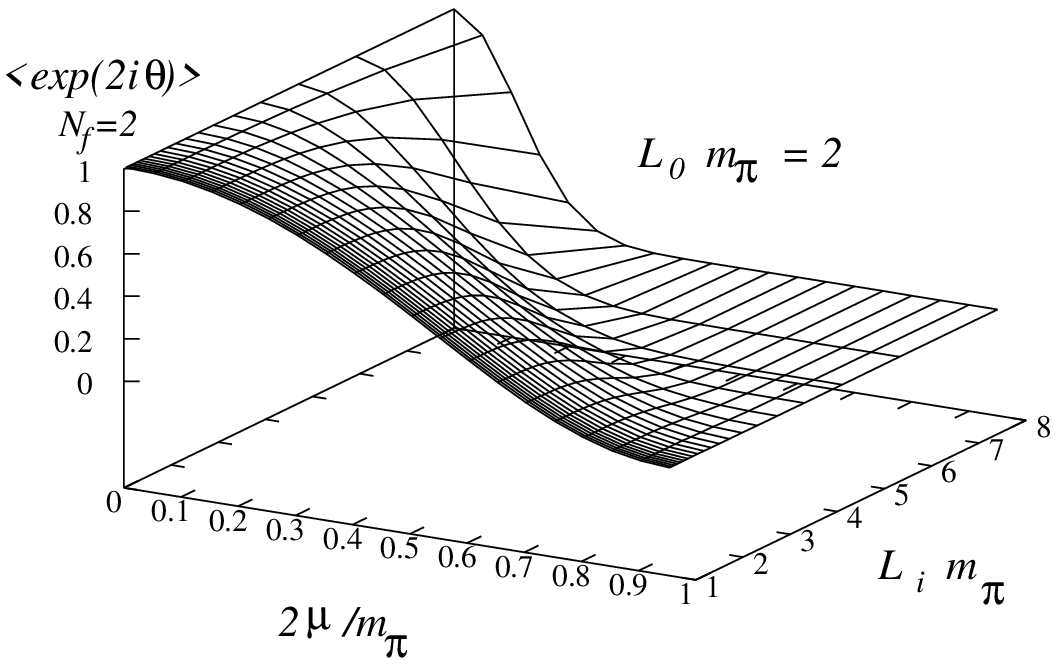,clip=,width=7cm}
  \vfill
  \epsfig{file=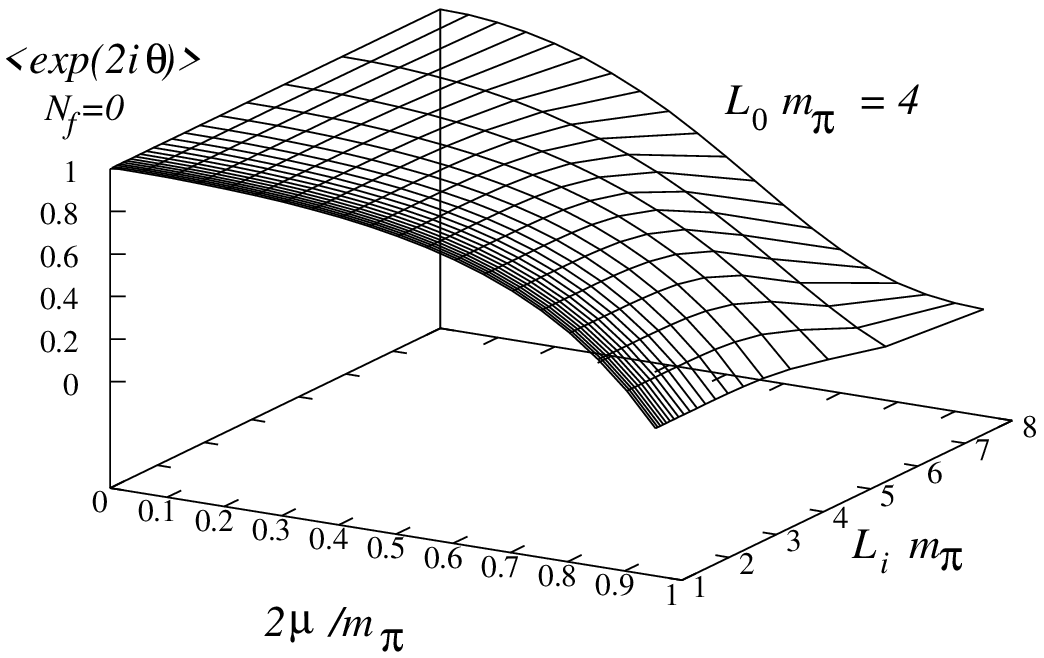,clip=,width=7cm}
 \epsfig{file=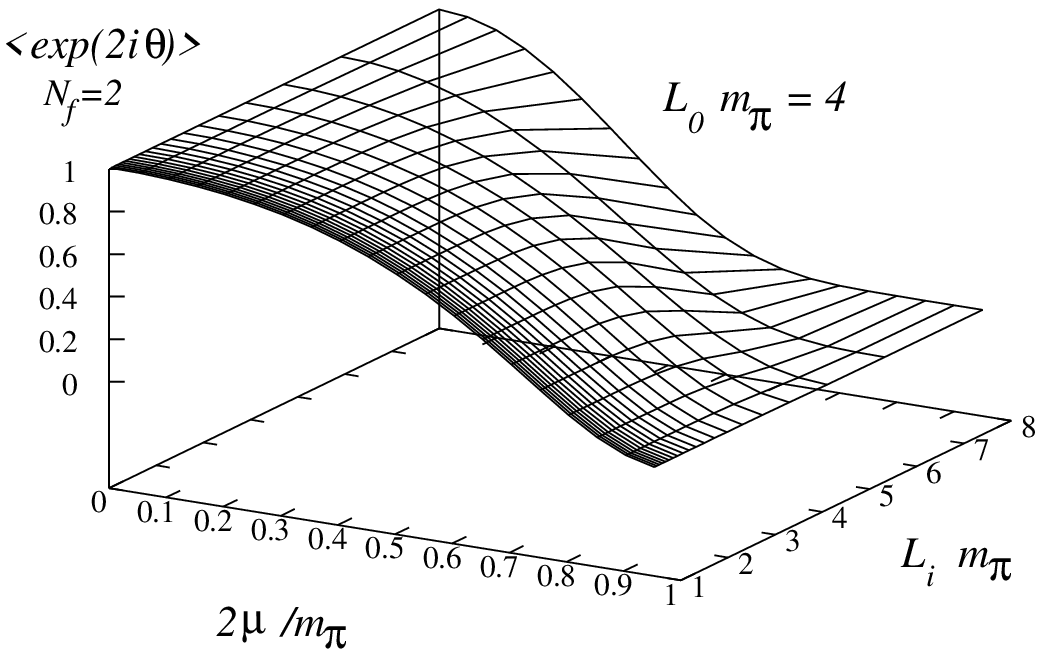,clip=,width=7cm}
    \caption{ \label{fig:mpiL0} The average phase factor goes to zero as the
    spatial extend of the box (in units of $1/m_\pi$) becomes large as compared
    to the temporal extend. The approach to the thermodynamic limit is 
    illustrated by varying $m_\pi L_i$ at fixed temperature $1/L_0$.
    {\bf Left:} The (phase) quenched average phase factor. 
{\bf Right:} The average phase
    factor with two dynamical flavors with the same mass.
    {\bf Top:} $m_\pi L_0=2$. {\bf Bottom:} $m_\pi L_0=4$.
    Note that the average phase factor starts dropping to zero 
    when $L_i$ exceeds $L_0$;  the spatial length  $L_i$
    needs to be larger than $L_0$ for the effect of the pion loop to
    be present in the average phase factor.}
\end{figure}

\begin{figure}[!h]
  \unitlength1.0cm
  \epsfig{file=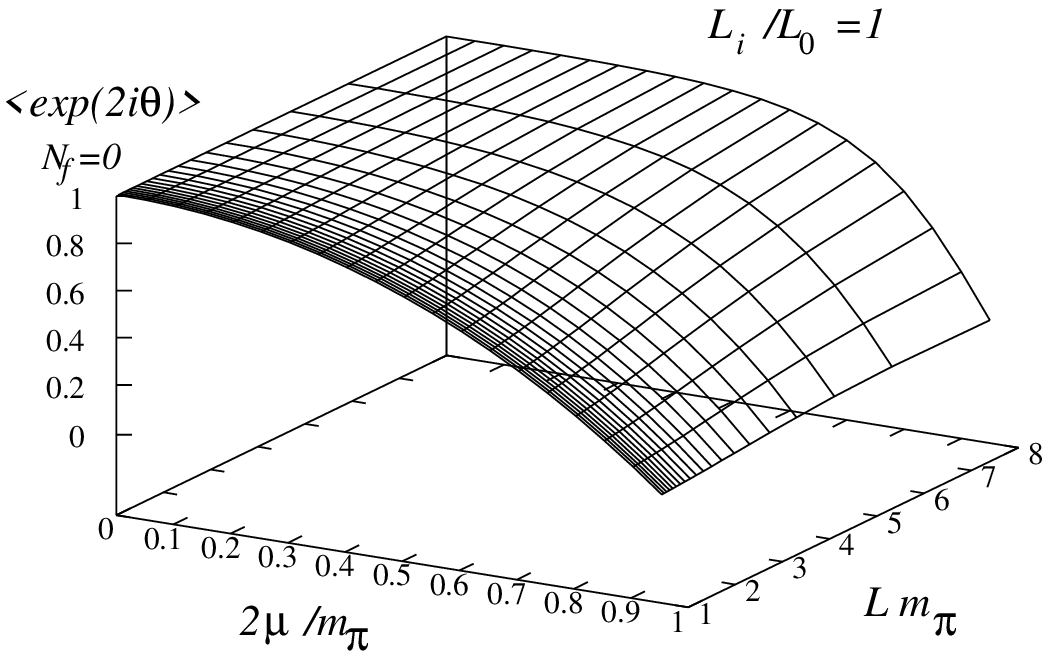,clip=,width=7cm}
 \epsfig{file=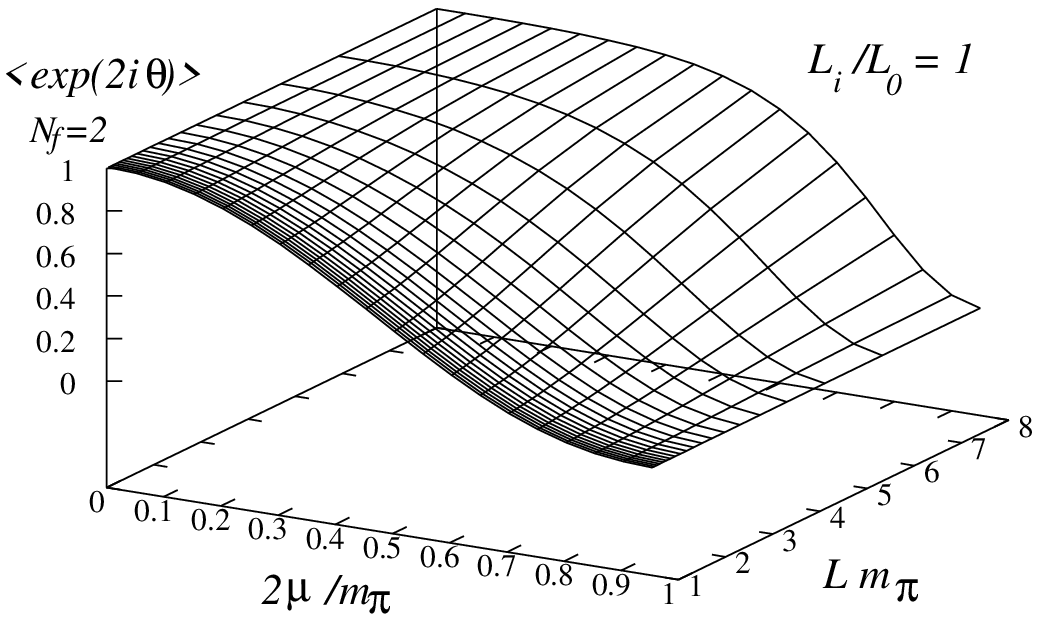,clip=,width=7cm}
  \vfill
  \epsfig{file=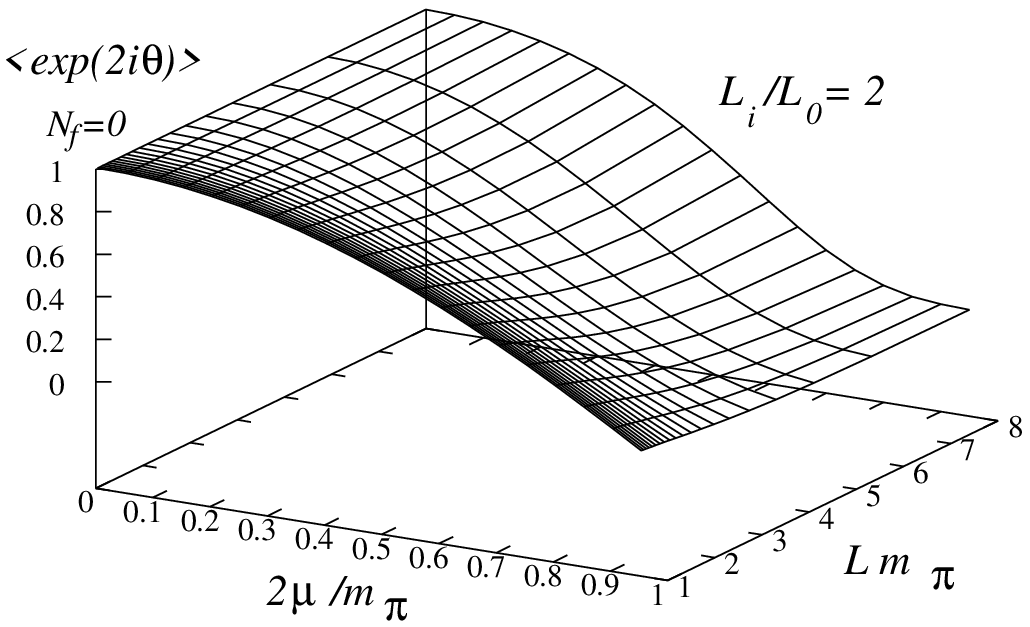,clip=,width=7cm}
 \epsfig{file=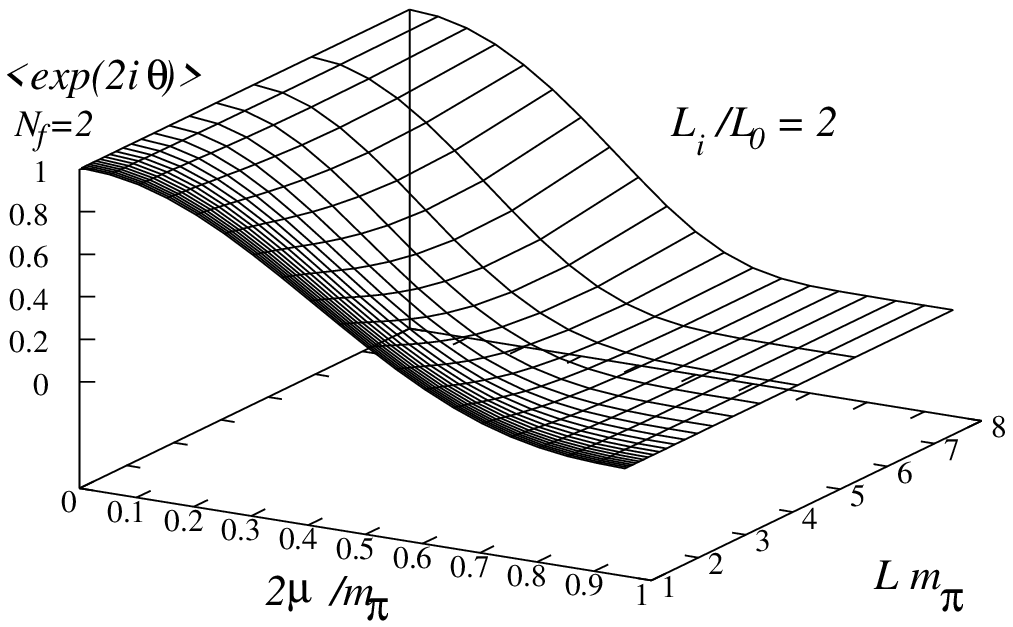,clip=,width=7cm}
  \caption{ \label{fig:LioverL01} The average phase factor when approaching
    the thermodynamic limit for fixed $L_i/L_0$. {\bf Left:} The (phase)
    quenched case. {\bf Right:} Two dynamical flavors. {\bf Top:} The temporal
    extent of the box is taken equal to its spatial extent, hence the
    $y$-axis is also the inverse temperature in units of $1/m_\pi$. 
      {\bf Bottom:} The asymmetry is now set to $L_i/L_0=2$. Hence
    the $Lm_\pi$ is also $\sqrt{8}L_0m_\pi$. In all cases, in the thermodynamic limit  the
    average phase factor approaches a step function which becomes zero
beyond    $2\mu=m_\pi$.}
\end{figure}

\end{widetext}

The one-loop result is valid for $L \Lambda_{QCD} \gg 1$, 
$m V \Sigma \gg 1$ and both $m_\pi \ll \Lambda_{QCD}$ and
$\mu \ll \Lambda_{QCD}$. Finally, the condition $\mu <m_\pi/2$
has to be satisfied. The reason is that for $\mu > m_\pi/2$, the
Goldstone fields have to be expanded about a rotated ground state. 
For $m_\pi^2L^2\ll1$ and $\mu^2L^2\ll1$ the dominant
contribution is given by the zero-momentum term and the one-loop result
reduces to (\ref{phase-hatmu-large}). A small $m_\pi L$ and small $\mu L$
expansion about this result was given in
\cite{phase-long}. The general one-loop result is also valid  when   
$m_\pi^2L^2$ and $\mu^2L^2$ are of
order or larger than 1. In the next section
 we study the one-loop result numerically.

\section{Numerical evaluation of the one-loop result}
\label{sec:num}
 
In this section we study the one-loop expression for the average phase factor
as a function of dimensionless combinations of the parameters. 
The average phase factor depends only on three dimensionless combinations:
\be
m_\pi L, \quad \mu L \quad {\rm and} \quad \frac {L_0}{L_i}.
\ee
As the one-loop expression is only valid for $\mu < m_\pi/2$ we will express
our results as a function of the ratio $\mu/(m_\pi/2)$.
We will
consider 3 different cases: 1)  {\sl The thermodynamic limit at fixed
  temperature} where $L_im_\pi$ becomes large while $L_0m_\pi$ 
remains 
fixed, 2)
{\sl The thermodynamic limit at low temperatures} where  $L_im_\pi$ grows for
fixed asymmetry $L_i/L_0,$ and 3) {\sl The low temperature limit 
in a finite box} where
$L_0m_\pi$ increases for fixed $L_im_\pi$.

\vspace{4mm}

\noi
{\bf Thermodynamic limit at fixed temperature:} 
The thermodynamic limit at fixed temperature is obtained by
letting the spatial extent of the box in units of the inverse 
pion mass go to
infinity while keeping the temporal extent also in units of $1/m_\pi$ fixed. 
In this limit the sign problem is acute for any nonzero value of the baryon
chemical potential, as indicated by a vanishing average phase factor. In
figure \ref{fig:mpiL0} we show the approach of
the average phase factor to
zero as $L_i$ increases. The two
left plots show the phase-quenched prediction (which, as discussed above, is
equivalent to the quenched prediction) while the two right  plots show
the case with two  dynamical flavors. Note that for $L_i<\L_0$ the 
average phase factor is dominated by the static pion modes
cf. (\ref{phase-hatmu-large}) \footnote{In the $p$-expansion, the 
microscopic variables $\mu^2F_\pi^2V$ and $m\Sigma V$ are large, and
consequently the leading term in the expansion matches the asymptotic 
limit of the microscopic expression provided that $m_\pi L\ll1$.}.

\vspace{4mm}

\noi
{\bf Thermodynamic limit with temperature going to zero:} As is common practice
in lattice QCD, 
we now fix the asymmetry of the box, $L_i/L_0$, and vary the size
of the box. A large box now automatically implies a low temperature. In this
thermodynamic limit the average phase factor approaches
 unity as long as the chemical
potential is less than half of the pion mass. For larger values of the
chemical potential the average phase factor is zero. In figure
\ref{fig:LioverL01} we show the approach to this
step function as a function of $L_i$. 
The two top panels are for a cubic box (quenched and unquenched) and the two
lower plots are for $L_i=2L_0$. The approach to the step function is slower
when the asymmetry is larger than unity.

\vspace{4mm}

\noi
{\bf Zero temperature in a finite box:} Here we consider the limit
where the spatial size of the box is fixed in units of $1/m_\pi$ as the
temperature (in units of $m_\pi$) is lowered. In the zero temperature
limit where the asymmetry $L_i/L_0\ll1$ the average phase factor again
approaches the step function $\theta(m_\pi-2\mu)$. This is not surprising given
that this behavior was also found in the exact solution of one-dimensional QCD
\cite{RV}.

\begin{figure}[!t]
  \unitlength1.0cm
  \epsfig{file=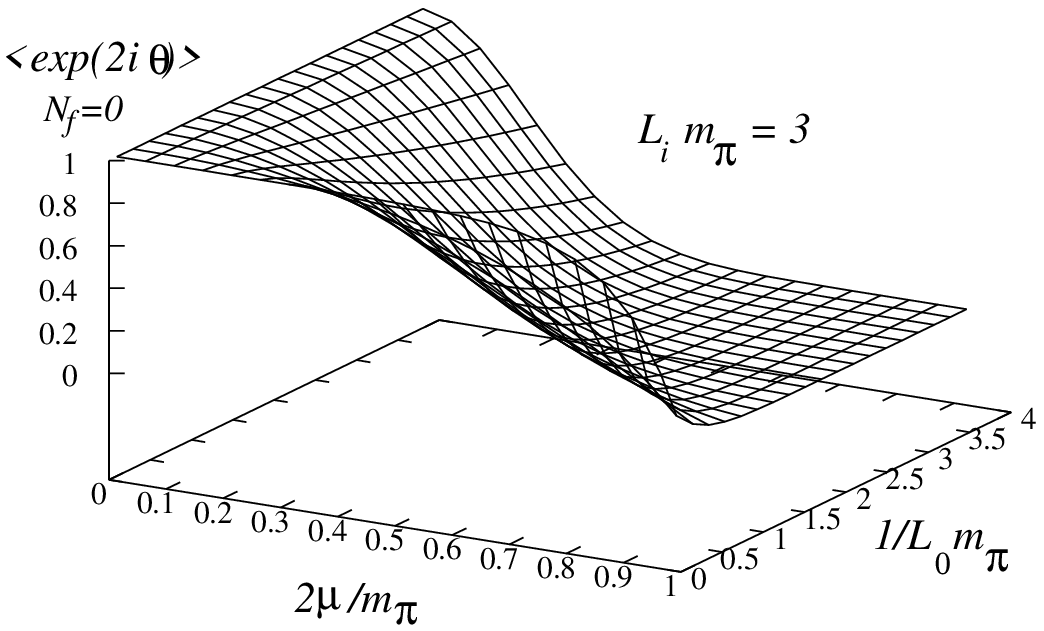,clip=,width=8cm}
  \epsfig{file=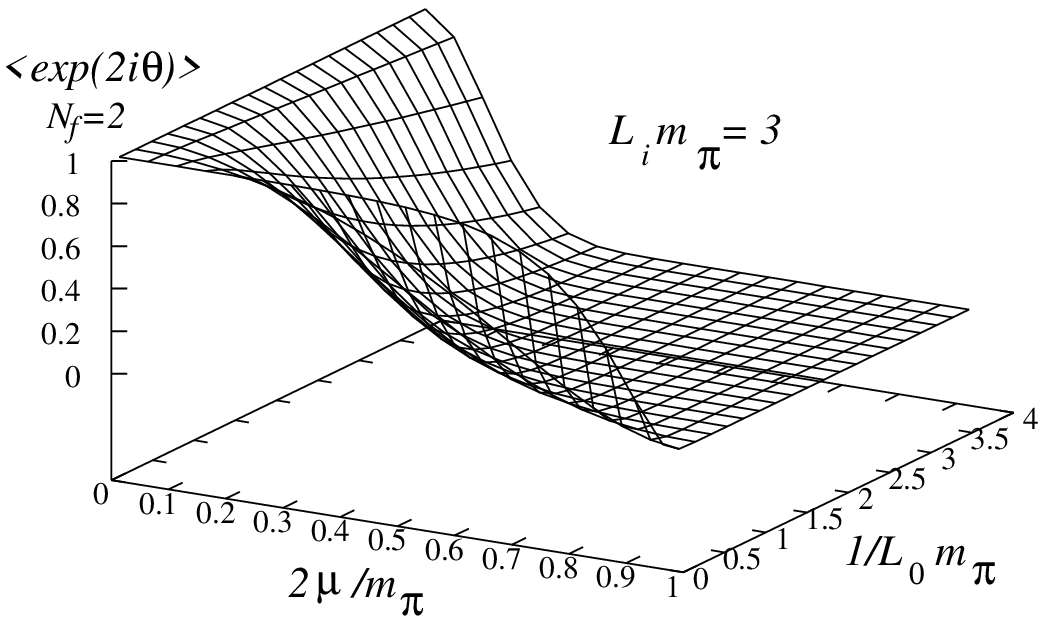,clip=,width=8cm}  
  \caption{\label{fig:mpiLi} The average phase factor for $m_\pi L_i=3$. It 
approaches zero with increasing temperature. {\bf
      Top:} The  (phase) quenched case. {\bf Bottom:} The $N_f=2$ case. 
Notice that the
    zero temperature limit is the step function $\theta(m_\pi-2\mu)$. In this
    limit the time direction is much longer than the spatial ones, and as
    expected, the step function found is consistent with the result from
    one-dimensional QCD \cite{RV}.} 
\end{figure}

In figure \ref{fig:mpiLi} we show how the average phase factor (quenched,
top, unquenched, bottom) behaves as a function of $\mu$ and $T$ in a finite
box.

\section{Comparison to lattice data by Allton {\it et al.}.}
\label{sec:lat}

The main limitations of lattice studies of QCD at nonzero chemical 
potential are the phase fluctuations of the fermion determinant and it
is natural to analyze them quantitatively
\cite{Gibbs,Toussaint,Schafer,deFL,NakamuraPhase,Ejiri1,Ejiri2,Allton1,schmidt,KimBen,Maria-LL*,Conradi}.
 In this section we compare the one-loop results obtained above to the lattice
data of Allton  {\it  et al.} \cite{Allton3}. 

In \cite{Allton3} the response of the QCD partition function 
to a baryon and isospin chemical potential was
measured at zero value of the chemical potential. The response to second
order in the baryon and isospin chemical potential was given in
terms of two numbers $c_2$ and $c_2^I$ respectively (see table 3.2 of
\cite{Allton3}).  They are related to the quark and isospin susceptibility 
at $\mu=0$ according to
\be
c_2 = \frac{\chi_q}{2T^2}, \qquad c_2^I = \frac{\chi_I}{2T^2}.
\ee

To second order in the chemical potential, the measured average phase
factor (in the phase quenched theory) is given by
\be
\langle e^{2i\theta}\rangle_{\rm lat} 
&=& e^{L_i^3 T \mu^2(c_2-c_2^I)}\\
&=& e^{(c_2-c_2^I)(L_i/L_0)^3(2\mu/m_\pi)^2(m_\pi/T_c)^2/(T/T_c)^2/4},\nn
\ee
where, in the second line, we have  expressed the result 
in terms of accessible dimensionless ratios.  
Note that the strength of the sign problem to lowest order in the Taylor
expansion only depends on the coefficient of the off-diagonal susceptibility 
$c_2^{ud}\equiv(c_2-c_2^I)/4$.

\begin{figure}[!t]
  \unitlength1.0cm
  \epsfig{file=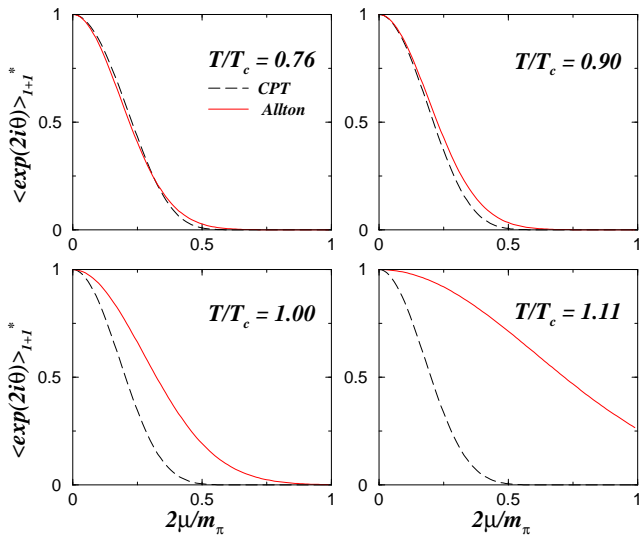,clip=,width=8.5cm}
\caption{\label{fig:Allton} The phase quenched average phase factor 
obtained using the data of \cite{Allton3} (solid curves) compared with the
result from chiral perturbation theory (dashed curves). The
temperatures are $T/T_c=0.76,0.90,1.00,1.11$ where $T_c$ is the critical
temperature at $\mu=0$. 
The agreement is good even close to
$T_c$.  Above the critical temperature the response to the baryon and isospin
chemical potential becomes alike as bound states of quarks
have melted. The prediction of chiral perturbation theory of course fails in
this region.}  
\end{figure}

The one-loop  
chiral perturbation theory result is obtained by matching the dimensions
of the box and the chemical potential in units of the inverse pion mass to 
those of \cite{Allton3}. To fix the scale we use the value 
$m_\pi/T_c=3.58$ from \cite{Karsch}. For temperatures below the critical
temperature, $T_c$, the  
agreement is very good, see figure \ref{fig:Allton}. Since chiral 
perturbation theory is not applicable in the chirally restored phase
the disagreement for $T> T_c$ is as expected.
Unfortunately, the isospin susceptibility was not calculated beyond 
second order in $\mu_I$ in \cite{Allton3} so that we cannot extract
the average phase factor to higher order.
The fourth order term is of particular interest because 
the analytical one-loop result, displayed in figure \ref{fig:Allton}, 
is well approximated by $\exp(-a\mu^2-b\mu^4)$ where $a$ and $b$ are positive
constants.   

Beyond the critical temperature 
the severity of the sign problem decreases significantly. 
In the high
temperature limit the difference from one of the average phase factor comes
from terms of order $g^4$ and higher as can be inferred from 
\cite{Reb,Vu}. Indeed,
the terms up to order $g^3$ are functions of $\sum_f \mu_f^2$ so that
the phase quenched and the unquenched partition function are
identical up to this order. In lattice simulations
by Allton {\it et al.}  the quark and isospin susceptibility are very
close for $T>T_c$. The physical implication is that bound states of
light quarks are absent beyond $T_c$.

%
%

\section{Finite versus infinite box}
\label{sec:resonance}

In this section we compare the result (\ref{g0mu}) for a  finite box  
with its thermodynamic limit which has
been used in the resonance gas model. 

In the resonance gas model it is usually assumed that $L_im_\pi\gg1$ 
such that the spatial momenta in the expression for $g_0$
can be integrated over instead of a discrete summation. This leads to the
standard expression
\be
\label{int-p}
g_0(\mu) =\frac{V m_\pi^2 T^2}{\pi^2} \sum_{n=1}^\infty \frac{K_2(\frac{m_\pi
    n}{T})}{n^2}\cosh(\frac{2\mu n}{T}).  
\ee
In addition, the resonance gas model includes heavier resonances. Among
others, it was applied \cite{edward-ch}
to the quark and isospin susceptibilities for
temperatures beyond $T_c$.

\begin{figure}[!t]
  \unitlength1.0cm
  \epsfig{file=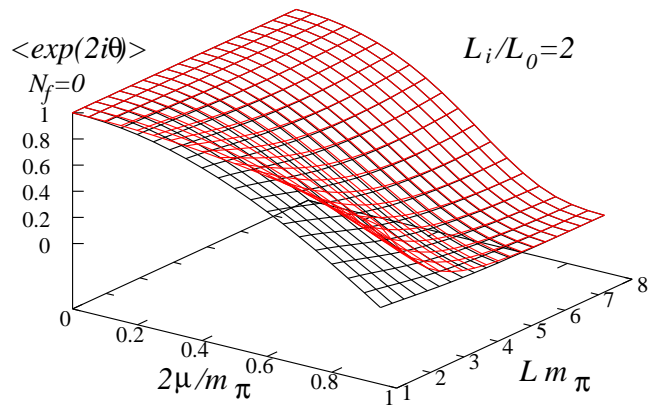,clip=,width=8.5cm}
  \caption{\label{fig:lim} The average phase factor for a finite box (lower
    surface) compared to the result in the thermodynamic limit at fixed
    $L_i/L_0$ (upper surface). Notice that finite size corrections become
    more important as $\mu$ increases.} 
\end{figure}

In Figure \ref{fig:lim} we compare the average phase factor in a box with
finite spatial length to the result obtained using (\ref{int-p}). 
We observe
that the validity of the standard expression (\ref{int-p}) depends on the
chemical potential. This is not surprising. After all the mass of the
charged Goldstone modes is $m_\pi\pm2\mu$ (recall that these Goldstone
modes are made out of a quark and a conjugate quark). As $\mu$ approaches
$m_\pi/2$ the lightest mode becomes massless which invalidates the
replacement the sum over momenta by an integral.

\section{The distribution of the phase}
\label{sec:theta-dist}

In addition to studying the average phase factor, it is natural
to also analyze the distribution function of the phase 
\cite{Toussaint,Ejiri:2007} itself. It is defined by
\be\label{rhoTh1} 
&&\rho_{N_f}(\theta)\equiv \langle\delta(\theta-\theta')\rangle_{N_f}\\
&&=\frac{\int dA |\det(D+\mu\gamma_0+m)|^{N_f} e^{iN_f\theta'}
  \delta(\theta-\theta') 
e^{-S_{\rm YM}}}{\int dA |\det(D+\mu\gamma_0+m)|^{N_f} e^{iN_f\theta'} 
e^{-S_{\rm YM}}} 
. \nn
\ee
The unquenched $\theta$-distribution can be written 
\be
\rho_{N_f}(\theta)
=e^{iN_f\theta}\rho_{N_f/2+N_f/2^*}(\theta)\frac{Z_{N_f/2+N_f/2^*}}{Z_{N_f}},
\label{rhoTh2} 
\ee
where the phase quenched $\theta$-distribution is defined as the average
\be
\rho_{N_f/2+N_f/2^*}(\theta)\equiv
\langle\delta(\theta-\theta')\rangle_{N_f/2+N_f/2^*}
\ee
with respect to the phase quenched partition function
\be
Z_{N_f/2+N_f/2^*}=\langle|\det(D+\mu\gamma_0+m)|^{N_f}\rangle.
\ee
This rewriting shows that the unquenched $\theta$-distribution is complex for
any nonzero value of the chemical potential. The complex nature of the
unquenched $\theta$-distribution resides entirely in $\exp(iN_f\theta)$ --
the other factors in (\ref{rhoTh2}) are real, positive and even. Despite
its simple form, the effect of the phase is drastic. Integrating over $\theta$
we find that 
\be
\int d\theta e^{iN_f\theta}\rho_{N_f/2+N_f/2^*}(\theta) =
\frac{Z_{N_f}}{Z_{N_f/2+N_f/2^*}}, 
\label{expSMALLint}
\ee
which becomes exponentially small ($\sim \exp(-V)$) in the thermodynamic limit
at fixed temperature.

In principle, one can extract the unquenched partition function from the phase
quenched one and the phase quenched $\theta$-distribution using
(\ref{expSMALLint}). Numerically it is, however, very difficult to handle the detailed
cancellations. It only works if the width of $\theta$-distribution 
is comparable to $2\pi/N_f$ so that 
the complex oscillations have little effect. 
Below we will compute the $\theta$-distribution in a finite box using chiral
perturbation theory and show 
that the width of the $\theta$-distribution is rather 
of order $\sqrt{V}$.

\subsection{$\theta$-distribution to one-loop in chiral perturbation theory}

The distribution function of the phase, $\rho_{N_f}(\theta)$, 
can be extracted from the moments of the phase factor. Just like 
the average phase factor the higher moments follow from the one-loop
computation    
\be
\langle e^{2ni\theta}\rangle_{N_f} 
&=& e^{n(N_f+n)(G_0(\mu=0)-G_0(\mu))}.
\label{exp2nith}
\ee
This holds for both both positive and negative integers $n$. 
Using the replica trick \cite{Edwards,replica}, one can analytically continue 
$n$ to noninteger values. For instance, 
\be
\langle e^{i\theta}\rangle_{N_f} = e^{\frac{1}{2}(N_f+\frac{1}{2})(G_0(\mu=0)-G_0(\mu))}.
\ee
We expect that the analytic continuation of $\langle \exp(2in\theta)\rangle$
in $n$ is valid when the quark mass is outside the support of the
eigenvalues -- as is always the case in this paper.

In~(\ref{exp2nith}) we evaluated the even Fourier components.
If we assume that the odd Fourrier coefficients are given by
the same expression 
we can simply express the delta-function in the definition of 
$\rho_{N_f}(\theta)$ into 
a sum over the moments. Introducing the shorthand   
\be
\Delta G_0\equiv G_0(\mu)-G_0(\mu=0)
\ee
we find
\be
\rho_{N_f}(\theta) 
&=& \frac 1{\pi} \sum_{n=-\infty}^\infty e^{-in\theta-(n/2)((n/2)+N_f) 
\Delta G_0} \nn \\
&=& \frac 1{\pi}e^{iN_f\theta +\frac 14 N_f^2 \Delta G_0} 
\sum_{n=-\infty}^\infty e^{-in\theta-n^2 \Delta G_0/4}\nn \\
&=& \frac 1{\pi}e^{iN_f\theta +\frac 14 N_f^2 \Delta G_0}
\vartheta_3(\theta/(2\pi), e^{-\Delta G_0/4}) .
\ee
By a Poisson resummation this can be rewritten as
\be
\rho_{N_f}(\theta) 
&=& \frac 1{\sqrt{\pi \Delta G_0}} e^{iN_f\theta +\frac 14 N_f^2 \Delta G_0}
\sum_n e^{-(\theta+2n\pi )^2/\Delta G_0},\nn \\
\ee
valid for a compact phase angle $\theta\in[-\pi,\pi]$.  
For a continuous phase angle, $\theta\in[-\infty,\infty]$ the distribution
function becomes a simple Gaussian:
\be
\rho_{N_f}(\theta) =  \frac{1}{\sqrt{\Delta G_0 \pi}}e^{(N_f/2)^2 \Delta G_0}
            e^{iN_f\theta-\theta^2 /\Delta G_0}.
\label{rhoth-CPT}
\ee
The quenched as well as the phase quenched average are given by
$\rho_{N_f=0}(\theta)$. Notice that result (\ref{rhoth-CPT}) is consistent with
the general form given in (\ref{rhoTh2}). Since $\Delta G_0$ is extensive, the
width of the $\theta$-distribution is of order $\sqrt{V}$ while its amplitude
increases exponentially with the volume. Along with the fact that the 
distribution is
normalized to one, this illustrates just how intricate the cancellations 
will be,
and therefore how tough the sign problem will be to handle numerically. 

With $\rho_{N_f}(\theta)$ at hand it is straightforward to compute also the
variance of the phase 
\be
\langle\theta^2\rangle_{N_f}-\langle\theta\rangle_{N_f}^2
= \frac{1}{2}\Delta G_0.
\ee
Note that the result is independent of $N_f$ even though 
$\langle\theta^2\rangle_{N_f}$ and $\langle\theta\rangle_{N_f}$ both 
depend on the
numbers of flavors. 
This suggest that the variance of the phase can be obtained from the
quenched theory.

The Gaussian form of the phase quenched $\theta$-distribution is in agreement
with the numerical result of Ejiri \cite{Ejiri:2007}. Notice, however, that an
error of order $1/\sqrt{V}$ in the numerical determination of the width of
the Gaussian will lead to an error of order one in the unquenched partition 
function, c.f. (\ref{expSMALLint}). Small non-Gaussian corrections can have a
similar dramatic effect. We expect that higher order terms in chiral
perturbation theory as well as effects from baryons will result in a 
non-Gaussian form.

\section{Conclusions}
\label{sec:concl}
The average phase factor of the fermion determinant has been computed and
examined to one-loop order in chiral perturbation theory 
for quark chemical potential
less than half of the pion mass. In the ordinary
thermodynamic limit at fixed temperature the average phase factor is zero 
for any nonzero value of the chemical potential. If the temperature is taken
to zero at fixed aspect ratio of the box,
the phase factor remains unity when the chemical potential is less than
$m_\pi/2$.  
This indicates that QCD at zero temperature has a mild sign 
problem for  $\mu<m_\pi/2$ in the thermodynamic limit. It would be of interest to study this
region on the lattice. In particular to examine if there is a spinodal line 
in this region. 

The one-loop prediction for the average phase factor is in agreement with
lattice data below the critical temperature. Below $T_c$ the  
one-loop prediction thus gives a direct way to estimate the strength of the
sign problem for a given lattice volume $V=L_i^3L_0$ and quark mass.  

The distribution of the phase itself was also derived from chiral
perturbation theory. Its simple Gaussian form is consistent with recent
lattice simulations \cite{Ejiri:2007}.

The critical isospin chemical potential beyond which pions Bose-Einstein 
condense is expected to depend on the temperature. The effect of this shift 
on the strength of the sign problem is not included in the present paper. 
In lattice simulations such effect has been  observed \cite{splitrev} and 
it would of great interest to extend the present work in this direction.

\vspace{4mm}

\noindent
{\sl Acknowledgments.} 
We wish to thank Misha Stephanov, Shinji Ejiri, Philippe de Forcrand, Simon
Hands and Maria Paola 
Lombardo for stimulating discussions as well as the Isaac Newton Institute
for Mathematical Sciences where this work was completed. This work was
supported  by U.S. DOE Grant No. DE-FG-88ER40388 (JV), the 
Carlsberg Foundation (KS), the Villum Kann Rassmussen Foundation (JV) and
the Danish National Bank (JV).

\end{document}